\documentclass[a4paper,11pt]{article}
\usepackage{here}
\usepackage{subfig}
\usepackage{graphicx}
\usepackage{braket}
\usepackage{amssymb, amsmath}
\usepackage[samesize]{cancel}
\usepackage{bm}
\usepackage{url}
\usepackage{jcappub}

\newcommand{\be}{\begin{eqnarray}}
\newcommand{\ee}{\end{eqnarray}}

\newcommand{\beq}{\begin{equation}}
\newcommand{\eeq}{\end{equation}}
\newcommand{\beqa}{\begin{eqnarray}}
\newcommand{\eeqa}{\end{eqnarray}}

\newcommand{\lmk}{\left(}
\newcommand{\rmk}{\right)}
\newcommand{\lkk}{\left[}
\newcommand{\rkk}{\right]}
\newcommand{\lnk}{\left\{}
\newcommand{\rnk}{\right\}}

\title{\boldmath Gravitational leptogenesis with  kination and gravitational reheating}

\author[a]{Kohei Kamada,}
\author[a,b]{Jun'ya Kume,}
\author[a]{Yusuke Yamada,}
\author[a,b,c]{\\ Jun'ichi Yokoyama}

\affiliation[a]{Department of Physics, Graduate School of Science,
The University of Tokyo,\\ Hongo 7-3-1
Bunkyo-ku, Tokyo 113-0033, Japan}
\affiliation[b]{Research Center for the Early Universe (RESCEU), Graduate School of Science,\\ The University of Tokyo, Hongo 7-3-1
Bunkyo-ku, Tokyo 113-0033, Japan}
\affiliation[c]{Kavli Institute for the Physics and Mathematics of the Universe (Kavli
 IPMU), UTIAS, WPI, The University of Tokyo, Kashiwa, Chiba, 277-8568, Japan}
\emailAdd{kohei.kamada@resceu.s.u-tokyo.ac.jp}
\emailAdd{kjun0107@resceu.s.u-tokyo.ac.jp}
\emailAdd{yamada@resceu.s.u-tokyo.ac.jp}
\emailAdd{yokoyama@resceu.s.u-tokyo.ac.jp}

\subheader{{\rm RESCEU-13/19}}

\abstract{We revisit the gravitational leptogenesis scenario in which the inflaton is coupled to gravity by the Chern-Simons term and the lepton asymmetry is generated through the gravitational anomaly in the lepton number current during inflation. We constrain the possible model parameter space by requiring the absence of ghost modes below the Planck scale, which would suggest the breakdown of the effective theory, and evaluate the net baryon asymmetry for various reheating processes. We find that the mechanism with these requirements is insufficient to explain the observed baryon asymmetry of the Universe for standard reheating scenarios. We show that, however, with the kination scenario realized in {\it e.g.} the $k$-
and G-inflation models, a sufficient baryon asymmetry can be generated within a feasible range of the model parameters.}

\begin{document}
\maketitle
\flushbottom
\section{Introduction}
The origin of matter-antimatter asymmetry of the Universe is one of the fundamental questions in cosmology. 
It has been quantitatively evaluated in terms of the baryon-to-entropy ratio from the 
observation of the Cosmic Microwave Background (CMB) as $n_B/s = (8.718 \pm 0.004)\times10^{-11}$~\cite{Aghanim:2018eyx}, 
which is consistent with the Big Bang Nucleosynthesis (BBN)~\cite{Cyburt:2015mya}. 
Since cosmic inflation dilutes all the preexisting baryon asymmetry, the asymmetry needs to be generated by a dynamics, so-called baryogenesis,  after (or during) inflation.

Among many  models of baryogenesis,  leptogenesis~\cite{Fukugita:1986hr} 
is one of the most noteworthy models, since we can obtain the baryon asymmetry simply by 
adding right-handed Majorana neutrinos to the Standard Model of particle physics, 
which also explain the non-vanishing neutrino mass for the observed neutrino oscillation.  
Here the lepton asymmetry is first generated through the lepton-number violating 
Majorana mass term and is converted to the baryon asymmetry by the electroweak
sphalerons~\cite{Kuzmin:1985mm}. 
In the ``vanilla'' leptogenesis scenario~\cite{Fukugita:1986hr}, 
the right-handed neutrinos are thermally produced 
and their decay generates the lepton asymmetry. 
Thus, the reheating temperature of the Universe $T_\mathrm{reh}$ must be higher than the lightest right-handed
neutrino mass $M_R$. 
In order to explain the present Universe, then, we have the lower bound of the right-handed 
neutrino mass as well as the reheating temperature, $T_\mathrm{reh} \gtrsim 10^9$ GeV
and $M_R \gtrsim 3 \times10^{9}$ GeV~\cite{Buchmuller:2004nz}.\footnote{Leptogenesis with lower right-handed 
neutrino mass is possible {\it e.g.} 
with a resonant process~\cite{Pilaftsis:2003gt} or 
the active-sterile neutrino oscillation~\cite{Akhmedov:1998qx}.}
However, if we generate lepton asymmetry with a different mechanism, 
we can be free from these constraints. 

One way to go beyond the vanilla leptogenesis is to consider a different  lepton-number 
violating interaction than the Majorana mass term. 
The gravitational leptogenesis proposed by Alexander, Peskin and Sheikh-Jabbari~\cite{AlvarezGaume:1983ig}
is a remarkable example. 
In that scenario, 
primordial lepton asymmetry is produced through the gravitational anomaly 
in the lepton current~\cite{AlvarezGaume:1983ig} 
during inflation through the generation of chiral gravitational waves induced by 
the gravitational Chern-Simons (gCS) coupling~\cite{Lue:1998mq,Choi:1999zy}. 
Since the gravitational chiral anomaly leads to the lepton current non-conservation
in the absence of the right-handed neutrinos, 
the lepton-number violating effect is effective only below the right-handed neutrino mass scale. 
The lepton asymmetry is generated directly in the Standard Model lepton sector and 
it does not require thermally-produced right-handed neutrinos. 
Therefore, in principle, we expect to have sufficient lepton asymmetry 
for a low reheating temperature
with a heavy right-handed neutrino mass. 
Unfortunately, quantitatively it turned out to be difficult to explain the present baryon asymmetry of the 
Universe~\cite{Alexander:2004us,Fischler:2007tj} unless 
an extremely large gCS coupling is 
realized~\cite{Alexander:2004us,Fischler:2007tj}.  
In order to overcome this difficulty, recently, gravitational leptogenesis from 
the chiral gravitational waves generated indirectly through the nontrivial gauge field configuration
in the pseudoscalar inflation was considered in the literature~\cite{Noorbala:2012fh,Maleknejad:2014wsa,Caldwell:2017chz,Adshead:2017znw}, which showed that 
the baryon asymmetry comparable to the observation can be generated in the SU(2) case (see also Ref.~\cite{Papageorgiou:2017yup} for the Abelian case). 
This is possible because the strength of the gauge Chern-Simons coupling can be 
naturally larger than the gravitational case. 
It is not clear, however, whether the original model with the inflaton-graviton coupling $f(\phi)R\tilde{R}$ can generate sufficient baryon asymmetry with the first option, 
namely, with an extremely large coupling. 
Indeed, it  turned out that the coupling may lead to the ghost instability of either the left- or right-polarized mode, which means we cannot predict anything on the evolution of the Universe
if its effect is too strong. 
To avoid such a too strong instability 
we may yield constraints on the model parameters, and
the coupling strength or the degree of CP violation turns out to be severely constrained~\cite{Alexander:2004wk,Lyth:2005jf}. 
One seriously needs to take into account such constraints in estimating the net amount of baryon asymmetry. 
We here note that the baryon asymmetry today highly depends on the reheating processes. 
We will show that, after taking all into account, the leptogenesis scenario fails 
in the ordinary reheating process with the 
inflaton's equation of state satisfying $w\leq 1/3$ after inflation until the completion of reheating 
due to the large entropy production. 

However, if entropy production at reheating is inefficient, the baryon-to-entropy ratio $n_B/s$ can be enhanced. 
We find that for the so-called "kination"~\cite{Joyce:1996cp} dominated reheating scenario~\cite{Spokoiny:1993kt}, where inflaton's equation of state is $w=1$, observed baryon asymmetry can be generated in the original gravitational leptogenesis model.\footnote{For baryogenesis in the kination domination scenario, see also Refs.~\cite{Joyce:1996cp, Bettoni:2018utf}.} Such a reheating scenario can be naturally realized in the $k$-inflation~\cite{ArmendarizPicon:1999rj} or G-inflation~\cite{Kobayashi:2010cm} with gravitational reheating~\cite{Ford:1986sy,Kunimitsu:2012xx,Figueroa:2016dsc,Nakama:2018gll,Dimopoulos:2018wfg,Hashiba:2018iff}.
We emphasize that the kination scenario as well as the kinetic term driven inflation models 
such as $k$-inflation or G-inflation are compatible with the gCS coupling. 
In the simplest case, the gCS term is proportional to $\phi R {\tilde R}$, 
which is shift symmetric in $\phi$. Such a shift symmetry explains the negligibly small potential of $\phi$ from the naturalness viewpoint.

In Sec.~\ref{RR}, we review the consistency of the chiral gravitational wave generation during inflation for models with the gCS term $f(\phi)R\tilde{R}$. 
We interpret this constraint in terms of the model parameter in the potential driven inflation 
as well as the $k$-inflation. 
We evaluate the generated chiral gravitational waves and the lepton number density  in Sec.~\ref{GL}. Starting from these results, we estimate the net baryon-to-entropy ratio 
for several reheating processes in Sec.~\ref{reheat}. We first show that it is impossible to obtain  sufficient baryon-to-entropy ratio for ordinary reheating processes. 
We also demonstrate that the kination scenario can be compatible with observation due to the inefficient entropy production.
Sec.~\ref{conc} is devoted for discussion.

\section{The consistency of the model with the gravitational Chern-Simons term during inflation} \label{RR}

Let us first review the chiral gravitational wave generation in the models with a gCS term following Ref.~\cite{Alexander:2004us}
and impose the constraints on the model parameters to avoid ghost instability 
as is done in Ref.~\cite{Alexander:2004wk}. 
The model consists of a real pseudo-scalar inflaton $\phi$ and its coupling to gCS term:
\beq
\begin{split}
S &= S_{{\rm E}-{\rm H}} + S_{\rm gCS} + S_{\phi} + S_{\rm matter}\\
  &= \int d^4x\sqrt{-g}\left[\frac{M_{\rm Pl}^2}{2}R+\frac{M_{\rm Pl}^2}{4}f(\phi)R\tilde{R} + \mathcal{L}_{\phi} + \mathcal{L}_{\rm matter}\right],\label{action}
\end{split}
\eeq
where
\beq
R\tilde{R} \equiv R_{\mu\nu\alpha\beta}\tilde{R}^{\mu\nu\alpha\beta} = \frac{1}{2}\frac{\epsilon^{\alpha\beta\gamma\delta}}{\sqrt{-g}}R_{\alpha\beta\rho\sigma}R_{\gamma\delta\mu\nu}g^{\mu\rho}g^{\nu\sigma}.
\eeq
Here we adopt the metric convention $g_{\mu\nu}=(-,+,+,+)$. 
$\epsilon^{\alpha\beta\gamma\delta}$ is the Levi-Civita tensor with $\epsilon^{0123} = +1$ and $M_{\rm Pl} \ (\sim 2.4\times 10^{18}{\rm GeV})$ is the reduced Planck mass. The existence of $f(\phi)R\tilde{R}$ term can be explained, {\it e.g.} by the Green-Schwarz mechanism in string theory~\cite{Green:1984sg}.

When $\phi$ field is dynamical during inflation and $f(\phi)$ varies 
as a classical background, CP invariance is broken in the gravity sector and hence
gravitational waves are polarized. 
In order to see it quantitatively, 
let us consider the following perturbed FRW metric,\beq
ds^2 = a(\eta)^2[-d\eta^2 + (\delta_{ij} + h_{ij})dx_idx_j].\label{metric}
\eeq
Here $\eta$ stands for the conformal time and $h_{ij}$ is the tensor perturbation 
in the transverse-traceless gauge.  
Expanding the first two terms of Eq.~(\ref{action}) up to the second order in $h_{ij}$, we find the quadratic action of the tensor perturbation as
\beq
\begin{split}
  S_{\rm GW}^{(2)} = \frac{M_{\rm Pl}^2}{8}\int d^4x & \left[ a^2(\eta)\lnk(h^i_{\ j})'(h^j_{\ i})' - (\partial_kh^i_{\ j})(\partial^kh^j_{\ i})\rnk \right. \\ 
    & \left.- f'\epsilon^{ijk}\lnk(h^q_{\ i})'(\partial_jh_{kq})' - (\partial^rh^q_{\ i})\partial_j\partial_rh_{kq}\rnk \right], \label{S2}
\end{split}
\eeq
where the prime represents the derivative with respect to the conformal time $\eta$. 
By moving to the Fourier space with 
circular polarization mode decomposition, it is simplified as follows~\cite{Alexander:2004us,Alexander:2004wk,Lyth:2005jf}. 
Let us expand the tensor perturbation $h_{ij}$ in the Fourier space as
\beq
h_{ij}(\eta,\textbf{x}) = \frac{1}{(2\pi)^{3/2}}\int d^3k\sum_{s = \mathrm{R,L}}p_{ij}^s(\bm{k})h_{\bm{k}}^s(\eta)e^{i\bm{k}\cdot \bm{x}}, \label{Fourier}
\eeq
where the circular polarization tensors are expressed in terms of the linear polarization tensors, 
$p_{ij}^{+, \times}({\bm k})$, as
\beq
\begin{split}
  p_{ij}^\mathrm{R} &\equiv (p^+_{ij} + ip_{ij}^{\times})/\sqrt{2},\\
  p_{ij}^\mathrm{L} &\equiv (p^+_{ij} - ip_{ij}^{\times})/\sqrt{2} = (p^R_{ij})^*. 
\end{split}  
\eeq
Here the linear polarization tensors are defined as 
\beq
\begin{split} 
p_{ij}^+ ({\bm k}) = ({\bm e_1}({\bm k}))_i  ({\bm e}_1({\bm k}))_j - ({\bm e_2}({\bm k}))_i  ({\bm e}_2({\bm k}))_j , \\
p_{ij}^\times ({\bm k}) = ({\bm e}_1({\bm k}))_i  ({\bm e}_2({\bm k}))_j + ({\bm e}_1({\bm k}))_j  ({\bm e}_2({\bm k}))_i, 
\end{split}
\eeq
where  $({\bm e}_1({\bm k}), {\bm e}_2({\bm k}), {\bm e}_3({\bm k}))$ with  ${\bm e}_3({\bm k}) \equiv {\bm k}/|{\bm k}|$ forms a right-handed, orthogonal triad of unit vectors and 
we assign ${\bm e}_s(-{\bm k})=-{\bm e}_s(-{\bm k}) \ (s = 1,2)$. Then the
circular polarization tensors satisfy following equations,
\beq
\begin{split}
  p_{ij}^\mathrm{R}({\bm k})p^{ij\mathrm{R}}({\bm k}) &= p^\mathrm{L}_{ij}({\bm k})p^{ij\mathrm{L}}({\bm k}) = 0,\\
  p_{ij}^\mathrm{R}({\bm k})p^{ij\mathrm{L}}({\bm k}) &= 2,\\
  k_p\epsilon^{mpj} p^A_{ij} ({\bm k}) &= -i\lambda^A k \ p^{m \  A}_{\ \ i}({\bm k}) \ (\text{for} \ A=\mathrm{L, R}), 
\end{split}
\eeq
where $\lambda^\mathrm{R} = +1, \lambda^\mathrm{L} = -1$. 
Note that they satisfy
$p^A_{ij}({\bm k}) = p^A_{ij}(-{\bm k}) \ (A = \mathrm{L,R})$
so that with the reality of the metric we have $h_{{\bm k}}^\mathrm{R} = (h_{-{\bm k}}^\mathrm{L})^*$.
Using these relations, the action (\ref{S2}) can be rewritten as
\beq
S_{\rm GW}^{(2)} = \frac{M_{\rm Pl}^2}{4} \int d\eta d^3k \sum_{A = L,R} a^2(\eta)\left[1 - \lambda^A k\frac{f'}{a^2(\eta)}\right](|(h^A_{{\bm k}})'|^2 - k^2|h^A_{{\bf k}}|^2).  \label{S2_f}
\eeq
By defining 
\beq
\mu_{\bm k}^A(\eta)  \equiv z_A (\eta, k) h_{\bm k}^A(\eta)
\eeq
with
\beq
z_A^2(\eta, k) = \frac{a^2M^2_{\rm Pl}}{2}\left(1 - \lambda^A k\frac{f'}{a^2}\right), 
\eeq
we obtain the mode equation 
\beq
 (\mu_{\bm{k}}^A)'' +\left(k^2 - \frac{z_A''}{z_A}\right)  (\mu_{\bm{k}}^A) = 0. 
\eeq
We can see that the  left- and right-handed modes are decoupled and obey different equations of motion.

Now let us discuss the conceptual problem in this quadratic action
and the mode equations 
as mentioned in the introduction. 
The kinetic coefficient of the tensor perturbation, $z_A^2(\eta, k)$, 
can cross zero and become negative for large $k$ modes 
with $\lambda_A f'>0$ at an early time when the scale factor is small 
with a monotonic evolution of $f(\eta)$. 
This means that there exist  ghost-like degrees of freedom in 
either  left- or  right-polarization modes. 
The apparent existence of ghost modes does not directly mean catastrophe, but 
they should not appear in low energy physics below the cut-off scale of the theory
since the presence of such ghost modes would spoil the validity of the effective theory. 
Then we can impose constrains on the behavior of the factor $z_A(\eta, k)$ 
so that the ghost modes would never appear below the cut-off scale of the theory~\cite{Alexander:2004wk}.  

Let us obtain constraints on the scalar field dynamics to avoid the problem. 
By neglecting the change in the Hubble parameter during inflation
so that $\eta = - 1/H a$, the factor $z_A(\eta, k)$ 
can be rewritten as
\beq 
z^2_A(\eta, k) = \frac{a^2M^2_{\rm Pl}}{2}\lmk1 - \lambda^A k \frac{\Theta}{8} \eta\rmk,\quad \Theta \equiv -8H\dot{f}(\phi), 
\eeq
where the dot denotes the derivative with respect to the physical time $t$ and 
$H={\dot a}/a$ is the Hubble parameter.  
Hereafter we focus on the case $\Theta>0$, but our discussion applies to the case with $\Theta<0$, too. 
By omitting the time dependence of $\Theta$,\footnote{As long as $\dot f(\phi)$ is dominated by the first linear term $\propto \dot\phi$, this approximation is valid in slow-roll inflation as well as k-inflation, in which $\dot\phi$ is effectively constant during inflation.} we can see that a left-handed mode with a given momentum $k$, $z_\mathrm{L}(\eta, k)$ is negative when 
\beq
\eta<\eta_{{\rm div}}(k) \equiv -\frac{8}{k\Theta}.
\eeq
In order for the system to be healthy, the physical wavenumber of the left-handed mode $k_{\rm phys}\equiv k/a$ should be larger than the  cut-off scale of the theory, $\Lambda$, at $\eta=\eta_\mathrm{div}(k)$ and gets below $\Lambda$ at a time $\eta>\eta_\mathrm{div}(k)$. 
This requirement can be rewritten as
\begin{equation}
k_\mathrm{phys}(\eta_{\rm div}) = \frac{k}{a(\eta_\mathrm{div})} = \frac{8H}{\Theta} \geq \Lambda.  
\end{equation}
We can see that the constraint is $k$-independent. 
Note that as long as $\Theta$  can be approximately constant, both $z_\mathrm{L}^2(\eta, k)$ and $k^{-1}_{\rm phys}(\eta)$ are monotonically increasing functions of $\eta$, 
and hence $z_\mathrm{L}^2(\eta, k)$ is positive for all the relevant modes. 
Strictly speaking, the cut-off scale of the theory is  $f(\phi)$-dependent,\footnote{Note that the cut-off scale or 
the strong-coupling scale of the fluctuations around the inflationary background 
is different from that in the vacuum~\cite{Bezrukov:2010jz,Germani:2011ua,Kamada:2015sca,Kunimitsu:2015faa}.}
but to give a conservative constraint we here take the cut-off scale to be the Planck scale, 
$\Lambda = M_\mathrm{Pl}$, 
since above the Planck scale regardless of $f(\phi)$ the system is strongly coupled 
in the pure gravity sector. 
From this requirement, $\Theta$, which parameterizes the degree of CP-violation, is constrained as~\cite{Alexander:2004wk}
\beq
 \Theta \lesssim \frac{H}{M_{\rm Pl}} \lesssim 10^{-5}.\label{constrain}
\eeq
Here we have used the upper bound of the Hubble parameter during inflation, 
$H\lesssim 10^{13}$~GeV,  
coming from the non-observation of the tensor modes in the CMB observations at Planck, 
$r < 0.11$~\cite{Aghanim:2018eyx} with $r$ being the tensor-to-scalar-ratio. 
Since this constraint does not depend on inflation models, we are able to constrain the gCS coupling independently of the details of the inflationary dynamics. 
In particular, it can be also applicable for the case when $\phi$ field is not the inflaton
but a spectator field since, in our discussion, the Hubble parameter is a free parameter independent of the $\phi$ field. 
Note that it has already been shown that the effect of chirality of the gravitational waves generated in this model is not observable in the CMB scale \cite{Alexander:2004wk} due to this constraint
(see also Ref.~\cite{Lyth:2005jf}). 

For concreteness and simplicity, let us consider the case where $f$ is linear with respect to $\phi$~\cite{Alexander:2004us}, 
\beq
f(\phi) = \frac{\mathcal{N}}{16\pi^2M_{\rm Pl}^2}\frac{\phi}{M_{\rm Pl}}. \label{fex}
\eeq
If we assume that the coupling is induced by the Green-Schwartz mechanism~\cite{Green:1984sg}, $\mathcal{N}$ is determined e.g., by the anomaly in original ten-dimensional theory and the volume of the compactified spaces.
As we will show the value of $\Theta$ relevant for the lepton asymmetry is that at the end of the inflation. In the case of potential-driven slow-roll inflation, $\Theta$ is determined by the Hubble parameter at the end of inflation $H_{\rm inf}$ and the first slow-roll parameter $\epsilon=(M^2_{\rm Pl}/2)\left(\partial_{\phi}V/V\right)^2$ as
\beq
\Theta_\mathrm{p} = \frac{\mathcal{N}}{2\pi^2}\left(\frac{H_\mathrm{inf}}{M_\mathrm{Pl}}\right)^2\sqrt{2\epsilon} = 7.3 \times 10^{-10}\lmk\frac{\mathcal{N}}{100}\rmk{\epsilon}^{1/2} \lmk\frac{H_{\rm inf}}{10^{-5}M_{\rm Pl}}\rmk^2.\label{theta}
\eeq
Thus for the parameters, $\mathcal{N} \sim 100$ and $H_{\rm inf} \lesssim 10^{-5}M_{\rm Pl}$, 
the constraint is easily satisfied. It also means that 
we can have the maximal gCS coupling $\Theta_\mathrm{max}=\frac{8H_\mathrm{inf}}{M_\mathrm{Pl}}$ that is allowed by the consistency of the theory only if ${\cal N}$ is as large as 
\beq
{\cal N} \sim 1.1\times 10^7  \epsilon^{-1/2} \lmk\frac{H_{\rm inf}}{10^{-5}M_{\rm Pl}}\rmk^{-1}. \label{Npot}
\eeq

If we go beyond the slow-roll inflation driven by potential, 
the situation can be changed.  
Here we take a simple realization of the  kinetically driven inflation or $k$-inflation~\cite{ArmendarizPicon:1999rj}, 
\beq
  \mathcal{L}(X,\phi) = \frac{K(\phi)X}{M_\mathrm{Pl}^2} + \frac{X^2}{M_\mathrm{Pl}^4},
\eeq
where $X \equiv -(1/2)(\nabla\phi)^2$. 
Inflation is realized when the function $K(\phi)$ satisfies following relations~\cite{ArmendarizPicon:1999rj}, 
\beq
\begin{split}
  3M_{\rm Pl}^2H^2_{\rm inf} \simeq \frac{1}{4}K^2(\phi), \\
  X = \frac{1}{2}\dot{\phi}^2 \simeq -\frac{M_{\rm Pl}^2}{2}K(\phi),
\end{split}
\eeq
so that the equation of state is $p \simeq -\rho $.
In this case, ${\dot \phi}$ is directly related to $H_\mathrm{inf}$ and 
the CP-violating parameter $\Theta$ defined in Eq.~\eqref{theta} is written as
\beq
\Theta_{\rm k} = (2\sqrt{3})^{\frac{1}{2}}\frac{\mathcal{N}}{2\pi^2}\left(\frac{H_{\rm inf}}{M_{\rm Pl}}\right)^{3/2} = 3.0\times10^{-7}\lmk\frac{\mathcal{N}}{100}\rmk\lmk\frac{H_{\rm inf}}{10^{-5}M_{\rm Pl}}\rmk^{3/2}.\label{theta_k}
\eeq
We see that $\Theta_{\rm k}$ is several orders of magnitudes larger than $\Theta_\mathrm{p}$, since in this expression, the power of the factor for $(H_{\rm inf}/M_{\rm Pl})$ is smaller than that in Eq.~\eqref{theta}. 
As a result, with the same $\mathcal{N}$ 
we have a stronger gCS coupling or the CP violation, and 
we obtain the maximal CP violation for smaller $\mathcal{N}$ than in the potential driven case shown in Eq.~\eqref{Npot}, 
\beq
{\mathcal N} \sim 10^4 \left( \frac{H_\mathrm{inf}}{10^{-5} M_\mathrm{Pl}}\right)^{-1/2}. 
\eeq
Note that for the parameters, $\mathcal{N} \sim 100$ and $H_{\rm inf} \lesssim 10^{-5}M_{\rm Pl}$, the degree of CP violation $\Theta$ still
satisfies the constraint from the consistency of the theory, Eq.~\eqref{constrain}. 
We emphasize that the gCS term with a linear coupling~\eqref{fex} 
is shift-symmetric in $\phi$, 
which forbids a potential for $\phi$ at a perturbative level.\footnote{Such shift symmetry would be broken by non-perturbative effects such as gravitational instantons.} Therefore $k$-inflation is compatible to the scenario 
better than potential-driven inflation models. 

\section{Chiral gravitational wave generation and gravitational leptogenesis}\label{GL}

Now we evaluate the generation of the gravitational waves and the lepton asymmetry 
through the gravitational anomaly in the lepton current. 
We further decompose the mode functions 
in terms of the creation and annihilation operators as 
\beq
\begin{split}
  \hat{\mu}_{{\bm k}}^\mathrm{R} &= u_{{\bm k}}^\mathrm{R}\hat{a}_{{\bm k}} + (u_{-{\bm k}}^\mathrm{L})^*\hat{b}^{\dagger}_{-{\bm k}},\\
  \hat{\mu}_{{\bm k}}^\mathrm{L} &= u_{{\bm k}}^\mathrm{L}\hat{b}_{{\bm k}} + (u_{-{\bm k}}^\mathrm{R})^*\hat{a}^{\dagger}_{-{\bm k}},\label{mode}
\end{split}
\eeq
where $\hat{a}_{{\bm k}}$ and $\hat{a}^{\dagger}_{{\bm k}}$ denote the annihilation and creation operator of the right-polarized mode with momentum ${\bm k}$, and $\hat{b}_{{\bm k}}$ and $\hat{b}^{\dagger}_{{\bm k}}$ denote those of the left-polarized mode, respectively. These operators satisfy commutation relations, $[\hat{a}_{{\bm k}}, \hat{a}^{\dagger}_{{\bm k'}} ] = [\hat{b}_{{\bm k}}, \hat{b}^{\dagger}_{{\bm k'}}] = \delta({\bm k} - {\bm k'})$.
Here we have used the hat to stress that they are quantum operators. 
For $f(\phi)$ linear in $\phi$~\eqref{fex}, in the slow-roll approximation 
$|{\ddot H}| \ll H^2$ and
$|{\ddot \phi}| \ll H |{\dot \phi}|$ (or more generally $|\Theta' \eta| \ll |\Theta|$), the mode equation is explicitly written as
\beq
(\mu_{\bm k}^A)'' + \left(k^2 - \frac{2}{\eta^2}-\frac{\lambda^A k \Theta/8}{(1-\lambda^A k \Theta \eta/8)\eta}+\frac{k^2 \Theta^2/256}{(1-\lambda^A k \Theta \eta/8)^2}\right)\mu_{\bm k}^A =0. 
\eeq
Here we take the slow-roll parameters to be zero.  
Due to the difference in $\lambda^A$  between the left- and right-polarization modes, 
the generated gravitational waves are circularly polarized. 
The generated gravitational wave spectrum can be estimated as follows.
At a later time when $|k \Theta \eta| \ll 1$, the mode equations are well approximated as
\beq
(\mu_{\bm k}^A)'' + \left(k^2 - \frac{2}{\eta^2}-\frac{\lambda^A k \Theta}{8\eta}-\frac{3k^2 \Theta^2}{256}\right)\mu_{\bm k}^A =0. 
\eeq
We have seen that at $k \Theta \eta  \sim -8$ the system is singular for one polarization mode,  and we cannot take the Bunch-Davies vacuum for the initial state as is in the usual case~\cite{Lyth:2005jf}. 
Here however we assume that the singular or non-linear effect around $\eta \sim -8/k \Theta$ 
does not affect the choice of the initial state at 
$\eta_\mathrm{i} > -8/k \Theta$  and the mode functions are described asymptotically by plane
waves, 
as discussed in Ref.~\cite{Alexander:2004wk}, so that the gravitational wave mode functions
at $ -8/k\Theta \ll \eta<0$ are given by 
\beq
\mu_\mathrm{k}^A(\eta) = \frac{1}{\sqrt{k} }e^{i k \eta_\mathrm{i}} \exp\left[-\frac{\lambda^A \pi \Theta}{32}\right] W_{\kappa, 3/2}\left(i \sqrt{4-\frac{3\Theta^2}{64}}k\eta \right), \quad \kappa \equiv \frac{i\lambda^A}{\sqrt{256/\Theta^2-3}}, \label{whitsol}
\eeq
where  $W_{\kappa,\mu}(z)$ is the Whittaker function. 
It has been argued that although due to the gCS term, 
the generated gravitational waves are modified as Eq.~\eqref{whitsol}
compared to the usual ones from inflation, 
taking into account the requirement Eq.~\eqref{constrain}
the modification is negligibly small~\cite{Alexander:2004wk}. 
However, it is true that the gravitational waves are circularly polarized,
and hence we can expect for the generation of lepton asymmetry as we will see
in the following. 

Let us examine lepton asymmetry generated in this model. 
Here lepton asymmetry is generated through the gravitational anomaly in the lepton current~\cite{AlvarezGaume:1983ig}, 
\beq
\nabla_{\mu}J_L^{\mu} = \frac{N_{\mathrm{R-L}}}{24(4\pi)^2}R\tilde{R},  \label{g-anomaly}
\eeq
where $N_\mathrm{R-L}$ is the difference in the number of species
of right- and left-handed leptons. 
Since the Standard Model of particle physics is chiral and the number of right- and left-handed Weyl fermions is different in the neutrino sector, 
the lepton number current $J_L^{\mu}$\footnote{Do not confuse the subscript $L$ for the leptons to the subscript L for the left-handed fermions.} is not conserved with $N_\mathrm{R-L} =3$, 
which corresponds to the number of flavor of neutrinos. 
For simplicity, let us assume that three heavy right-handed Majorana neutrinos exist with 
(degenerate) Majorana mass $\mu$. Then the gravitational anomaly is cancelled 
at the energy scales higher than $\mu$, 
and the Majorana mass $\mu$ acts as the cut-off scale of the gravitational anomaly. 

We have seen that with the gCS coupling, circularly polarized gravitational waves 
are generated when the scalar field $\phi$ is dynamical, 
which means that the gravitational Chern-Pontryagin density, $R {\tilde R}$ 
acquires a non-vanishing expectation value $\langle R {\tilde R} \rangle$. 
Then through the gravitational anomaly in the lepton current~\eqref{g-anomaly}, 
lepton asymmetry is generated. 
By expressing the Chern-Pontryagin density in terms of the divergence of the Chern-Simons
current, $R{\tilde R} = \nabla_\mu K^\mu$, 
this process can be understood as the improved conservation law of the 
lepton current, 
\beq
\nabla_{\mu} \left(J_L^{\mu} -  \frac{N_\mathrm{R-L}}{24(4\pi)^2} K^\mu \right)= 0. 
\eeq
Assuming that there are no asymmetry in both the lepton number and gravitational waves 
at the beginning, by taking the spatial average, 
the lepton number asymmetry at the end of inflation $\eta_\mathrm{f}$ is expressed as
\beq
{\bar n_L}(\eta_\mathrm{f}) = \frac{N_\mathrm{R-L}}{24(4\pi)^2} a_\mathrm{f} \left.{\bar K^0}(\eta_\mathrm{f})\right|_{k/a_\mathrm{f}<\mu}, 
\eeq
with the same analogy to the case of the cogeneration of gauge fields and fermions 
in the pseudoscalar inflation with the $\phi F {\tilde F}$ coupling~\cite{Domcke:2018eki,Domcke:2018gfr,Domcke:2019mnd}.  
Here $n_L \equiv a J_L^0$ is the lepton asymmetry density and the subscript f 
represents that the variable is evaluated at the inflation end.
The scale factor $a$ is introduced to make the quantity physical but not comoving one.
Note that we have taken into account the cut-off scale of the anomaly equations. 

With the metric \eqref{metric}, the Chern-Pontryagin density $R\tilde{R}$ can be written as
\beq
a^4R\tilde{R} = \partial_{\eta}\lkk\frac{1}{2}\epsilon^{ijk}(-\partial_lh_{jm}\partial_m\partial_ih_{kl} + \partial_lh_{jm}\partial_l\partial_ih_{km} - h'_{jl}\partial_ih'_{lk})\rkk + \partial_i(\cdots).\label{RR_dual}
\eeq
Here the ellipses part is the spatial component of the Chern-Simons current and 
we do not write it explicitly. 
The zero-th component of the Chern-Simons current is given by 
\beq 
a^4 K^0 = \frac{1}{2}\epsilon^{ijk}(-\partial_lh_{jm}\partial_m\partial_ih_{kl} + \partial_lh_{jm}\partial_l\partial_ih_{km} - h'_{jl}\partial_ih'_{lk}) . 
\eeq
Then the expectation value of the spatial average of the zero-th component of the 
Chern-Simons current or the lepton asymmetry is given by 
\begin{align}
\langle {\bar n_L} (\eta_\mathrm{f})\rangle &= \frac{N_\mathrm{R-L}}{24(4\pi)^2}a_\mathrm{f} \left.\langle {\bar K^0}(\eta_\mathrm{f})\rangle \right|_{k/a_\mathrm{f}<\mu}\notag \\
& = \frac{N_\mathrm{R-L}}{48 (4\pi)^2Va_\mathrm{f}^{3}}  \int_{k/a_\mathrm{f}<\mu} d^3 x (\epsilon^{ijk}\langle (\partial_lh_{jm}\partial_l\partial_ih_{km} -h'_{jl}\partial_ih'_{lk})\rangle)|_{\eta_\mathrm{f}},
\end{align}
where $V = \int d^3 x$ is the spatial volume.  

By expanding $h_{ij}$ in Fourier space as Eq.~\eqref{Fourier} and replacing $h^A_{{\bf k}}(\eta_\mathrm{f})$ to the quantum operator $\hat{\mu}^A_{{\bf k}}(\eta_\mathrm{f})$, we obtain 
\beq
\braket{{\bar n_L}(\eta_\mathrm{f})} = \frac{N_\mathrm{R-L}}{384\pi^2a_\mathrm{f}^3}\int_{k/a_\mathrm{f}<\mu} \frac{d^3k}{(2\pi)^3} {\left[k^3\left(|h^\mathrm{R}_{{\bf k}}(\eta_\mathrm{f})|^2-|h^\mathrm{L}_{{\bf k}}(\eta_\mathrm{f})|^2\right) - k\left(|(h^\mathrm{R}_{{\bf k}}(\eta_\mathrm{f}))'|^2 - |(h^\mathrm{L}_{{\bf k}}(\eta_\mathrm{f}))'|^2\right) \right]}. \label{lep_ev}
\eeq

By substituting Eq. (\ref{whitsol}) to Eq. (\ref{lep_ev}), one finds the lepton number density 
up to the first order in $\Theta$~\cite{Alexander:2004us},\footnote{The numerical factor of our expression is different from that in \cite{Alexander:2004us} mainly due to the different definition of $\Theta$ and the factor of 24 in~\eqref{g-anomaly}.}
\beq
\braket{{\bar n}_L} = -\frac{1}{2048\pi^4}\left(\frac{H_\mathrm{f}}{M_{\rm Pl}}\right)^2\Theta H_\mathrm{f}^3\left(\frac{\mu}{H_\mathrm{f}}\right)^4, \label{lep}
\eeq
where $H_\mathrm{f}$ is the Hubble parameter at the end of inflation $\eta=\eta_\mathrm{f}$
and we have used $N_\mathrm{R-L}=3$. 
Here the value of $\Theta$ should be taken to be the one at the end of inflation,
since the lepton asymmetry is determined by the properties at the boundary or the inflation end. 
The amplitude of 
$\Theta$ exhibits a damped oscillation for the reheating scenario with inflaton oscillation 
and smooth decay for the kination scenario, 
and hence we do not expect for a substantial change of lepton asymmetry during reheating stage.

One can see that the lepton asymmetry is highly sensitive to the cut-off scale $\mu$. 
This is because the Chern-Pontryagin density is UV divergent. 
In order to evaluate the gravitational waves we need to know the physics
beyond the cut-off scale of the theory $\Lambda$, which we identified to be the Planck scale. 
However, as long as the cut-off scale for the gravitational anomaly satisfies 
$\mu \ll \Lambda\sim M_\mathrm{Pl}$, the relevant wave number of the gravitational waves
for the leptogenesis 
is well below the cut-off scale of the theory, and hence we believe that the estimate 
for the lepton asymmetry is valid. 

\section{The net baryon asymmetry $n_B/s$ in the gravitational reheating scenario}\label{reheat}
We have obtained the generated lepton number density in the previous section.  
Let us evaluate the baryon-to-entropy ratio $n_B/s$ today. 
Here we note that 
the baryon number density and the entropy density today highly depend on the thermal history after inflation before the completion of reheating, which has huge uncertainties even if we fix 
the dynamics during inflation. 
By parameterizing it with the (effective) equation of state during reheating is $p = w\rho$
with $w$ being a constant, we obtain lepton-to-entropy ratio $n_L/s$ at the completion 
of reheating in terms of the reheating temperature $T_{\rm reh}$ as
\begin{align}
\frac{n_L(\eta_{\rm reh})}{s_{\rm reh}} &= \frac{n_L(\eta_\mathrm{f})}{s_{\rm reh}}\epsilon_{\rm w.o}\left(\frac{a_\mathrm{f}}{a_{\rm reh}}\right)^3 \notag \\
&= -\frac{45}{4096\pi^6}\epsilon_{\rm w.o}\lmk\frac{\pi^2}{90}\rmk^{\frac{1}{1+w}}g_*^{\frac{1}{1+w}} g_{*s}^{-1}\left(\frac{T_{\rm reh}}{M_\mathrm{Pl}}\right)^{\frac{1-3w}{1+w}}\left(\frac{H_\mathrm{f}}{M_\mathrm{Pl}}\right)^{-\frac{1-w}{1+w}}\left(\frac{\mu}{M_\mathrm{Pl}}\right)^4\Theta,
\end{align}
where $g_{*}$ and $g_{*s}$ are the effective number of relativistic degrees of freedom 
for the energy density and entropy at the reheating, respectively, and we take $g_*\simeq g_{*s} \sim 100$ for standard model as a reference value.
Here we have taken into account the washout effect due to the lepton-number violating Majorana neutrino mass with the washout factor $\epsilon_{\rm w.o}$. 
It has been noticed that in the regime $\mu \gg H_\mathrm{f}$ the washout factor is 
not exponentially suppressed but at most 0.1~\cite{Adshead:2017znw} due to the approximate 
conservation of the right-handed electron number at $T \gg$ 80 TeV~\cite{Campbell:1992jd}.

The lepton asymmetry is transferred to the baryon asymmetry via the electroweak sphaleron~\cite{Kuzmin:1985mm} before the electroweak symmetry breaking and the baryon-to-entropy ratio is conserved 
afterwards. 
Thus if the reheating temperature is higher than the electroweak scale, 
the resulting baryon asymmetry is~\cite{Harvey:1990qw}
\beq
\frac{n_B}{s} = -\frac{28}{79}\epsilon_{\rm w.o}\frac{n_L}{s} = \frac{315}{80896\pi^6}\epsilon_{\rm w.o}\lmk\frac{\pi^2}{90}\rmk^{\frac{1}{1+w}}g_*^{-\frac{w}{1+w}}\left(\frac{T_{\rm reh}}{M_\mathrm{Pl}}\right)^{\frac{1-3w}{1+w}}\left(\frac{H_\mathrm{f}}{M_\mathrm{Pl}}\right)^{-\frac{1-w}{1+w}}\left(\frac{\mu}{M_\mathrm{Pl}}\right)^4\Theta, \label{baryon}
\eeq
where we have assumed that $g_{*}=g_{*s}$.

For $w \leq 1/3$, we can see that from Eq.~\eqref{baryon} the baryon asymmetry becomes smaller as $T_{\rm reh}$ decreases
because in this case 
the Universe is dominated by non-extraordinary fluids which includes relativistic and non-relativistic ones and lower reheating temperature 
means larger entropy production. 
Thus in the case of instantaneous reheating, $T_\mathrm{reh} = (90/\pi^2 g_*)^{1/4} (H^2_\mathrm{f} M^2_\mathrm{Pl})^{1/4}$, the maximal net baryon-to-entropy ratio 
is obtained. 
In this case, Eq. (\ref{baryon}) reads
\beq
\frac{n_B}{s} = 1.5\times 10^{-19} \epsilon_{\rm w.o} \lmk\frac{g_*}{100}\rmk^{-\frac14}\lmk \frac{H_\mathrm{f}}{10^{13}\mathrm{GeV}} \rmk^{1/2} \lmk\frac{\mu}{10^{16} \mathrm{GeV}}\rmk^4\lmk\frac{\Theta}{H_\mathrm{f}/M_\mathrm{Pl}}\rmk. 
\eeq
Note that for larger Majorana neutrino mass $\mu > 10^{16}$ GeV 
the perturbativity of the neutrino Yukawa interaction breaks down for the neutrino mass 
compatible with the observed nonzero neutrino oscillation~\cite{Abazajian:2012ys}, 
and hence the factor $(\mu/10^{16} \mathrm{GeV})$ is required to be less than the unity. 
With the constraint on the Hubble parameter during inflation, $H_\mathrm{f}<10^{13}$ GeV, 
coming from  the constraint on the tensor-to-scalar ratio in the CMB, $r<0.11$, as well as 
the constraint on the strength of the gCS coupling $\Theta$ (Eq.~\eqref{constrain}), 
we conclude that the baryon-to-entropy ratio in this scenario is at most $n_B/s\simeq 1.5 \times
10^{-19}$, 
which is far smaller than the observed value, and this mechanism 
cannot be responsible for the present baryon asymmetry of the Universe. 

For $w > 1/3$, however, the situation can be dramatically changed. 
Since the energy density of the inflaton decays faster than radiation, 
entropy production at reheating is less efficient than the case with instantaneous reheating. 
As a result, $n_B/s$ is larger when the reheating temperature is lower.
Thus the possibility to realize the baryon asymmetry comparable to the observed value opens up at a low reheating temperature. 
Note that $w=1$ is realized, {\it e.g.},  in the case of kination~\cite{Spokoiny:1993kt}
where the kinetic energy of the scalar field without a potential energy  dominates the Universe. 
In the case 
with $w = 1$, 
Eq.~\eqref{baryon} reads
\beq
\frac{n_B}{s} = 9.7 \times 10^{-11} \epsilon_{\rm w.o}\lmk\frac{g_*}{100}\rmk^{-1/2}\lmk\frac{\mu}{10^{16}\mathrm{GeV}}\rmk^4\lmk\frac{\Theta}{10^{-5}}\rmk\lmk\frac{T_{\rm reh}}{10^7\mathrm{GeV}}\rmk^{-1}.
\eeq
Thus due to the insufficient entropy production in the kination scenario, 
the present baryon asymmetry  $n_B/s \sim 8.7 \times 10^{-11}$ can be explained for a reasonable
reheating temperature $T_\mathrm{reh} \simeq 10^{7}$ GeV. 
Note that the resultant baryon asymmetry does not depend on the Hubble parameter during inflation. 
As a result, we can take $H_\mathrm{f} \sim 10^{-5} M_\mathrm{Pl}$ 
so that $\Theta\simeq 10^{-5}$ is allowed by satisfying the consistency condition (Eq.~\eqref{constrain}), {\it e.g.}, with ${\cal N} \sim 10^4$ for the gCS coupling Eq.~\eqref{fex}.

Before concluding, let us comment on the issue if the reheating can complete in the kination scenario. 
Since  inflaton continues to roll and its coherent oscillation does not occur, 
it is impossible to reheat the Universe by the decay of inflaton. 
The standard reheating mechanism in the kination scenario is 
the gravitational reheating~\cite{Ford:1986sy,Kunimitsu:2012xx,Figueroa:2016dsc,Nakama:2018gll,Dimopoulos:2018wfg,Hashiba:2018iff}, 
in which the particles are produced gravitationally by the change of the background spacetime~\cite{Parker:1969au}, and eventually dominate the energy density of the Universe. 
However, it is not clear if the relativistic particles generated by this process
are successfully thermalized up to the Standard Model sector 
well before the BBN or the electroweak symmetry breaking 
(required for the successful leptogenesis). 
Note that the gauge fields and (massless) fermions are conformally coupled and hence 
they are not directly produced by gravitational reheating. 

Let us show a toy model that is free from the concern in the above. 
Suppose a massive spectator field conformally coupled to gravity, 
\beq
S_{\chi} = \int d^4x \sqrt{-g}\lmk -\frac{1}{2}g^{\mu\nu}\nabla_{\mu}\chi\nabla_{\nu}\chi -\frac{1}{12}\chi^2R - \frac{1}{2}m^2\chi^2\rmk, 
\eeq
with $m \lesssim H_\mathrm{f}$, which has a tiny coupling to the Standard Model sector. 
At the transition between the epoch of inflation and kination, 
$\chi$ fields are produced gravitationally and 
acquire the energy density $\rho_\chi \sim 2\times 10^{-4}e^{-4m\Delta t} m^2 H_\mathrm{f}^2$, where $\Delta t\sim H_f^{-1}$ is the transition time for de Sitter to kination regime~\cite{Hashiba:2018iff}. 
They behave as non-relativistic particles just after the onset of kination 
since their typical wave number is the mass scale~\cite{Hashiba:2018iff}, 
and hence they can eventually dominate the energy density of the Universe. 
Note that while the energy density of the Universe during kination decays with $\rho \propto 
a^{-6}$ whereas those of the $\chi$ field decays with $\rho_\chi \propto a^{-3}$. 
We assume that through the tiny coupling to the Standard Model sector, {\it e.g.}, 
$(\lambda/M_{Pl})\chi F_{\mu\nu}F^{\mu\nu}$, $\chi$ fields decay into 
the Standard Model particles with a decay width $\Gamma = \alpha m$.  
If the decay happens during kination 
before the $\chi$ fields dominate the energy density of the Universe, 
which is realized for $\alpha \gtrsim 2\times 10^{-4}e^{-4m\Delta t} H_\mathrm{f} m/M_\mathrm{Pl}^2$, 
the relativistic decay product finally dominates the energy density of the Universe. 
Since we assume that the decay products belong to the Standard Model, 
these particles are smoothly thermalized and hence the reheating 
of the Universe is completed until when they dominate the energy density of the Universe. 
We can easily check that for $m\sim H_\mathrm{f} \simeq 10^{13}$ GeV and $\alpha \simeq 2\times 10^{-4}e^{-4m\Delta t}H_\mathrm{f} m/M_\mathrm{Pl}^2$ (or $\lambda \simeq (8\pi/10^3)^{1/2} ( H_\mathrm{f}/m)^{1/2} \sim 0.1$), the Universe is dominated by the relativistic particles
at the Hubble parameter $H \sim 3 \times 10^{-2}$ GeV, 
which corresponds to the reheating temperature $T_\mathrm{reh} \simeq 1.6e^{-2m\Delta t}\times 10^8$ GeV. When $m\sim H_f\sim \Delta t^{-1}$, the exponential factor can be $\mathcal{O}(0.1)$, which realizes the reheating temperature $T\sim 10^{7}$ GeV.
Thus, with an efficient reheating mechanism shown above, 
the gravitational leptogenesis with kination is a plausible model to explain the 
present baryon asymmetry of the Universe. 
Since the gravitational leptogenesis works for relatively heavy Majorana neutrino mass case
$M_R \sim 10^{16}$ GeV, it is complementary to the case studied in Ref.~\cite{Hashiba:2019mzm} 
in the point of view of the gravitational reheating scenario.

\section{Discussion} \label{conc}
The gravitational leptogenesis scenario~\cite{Alexander:2004us} is a novel mechanism 
to produce the lepton asymmetry from the chiral gravitational waves generated during inflation 
with a gCS coupling  through the gravitational anomaly in the lepton current. 
In this paper, we have reinvestigated whether it can healthily lead to baryon asymmetry comparable to the observed value. 
As is already pointed out in Ref.~\cite{Alexander:2004wk}, the degree of CP-violation in this model is strongly constrained to avoid the ghost instability, regardless of the inflation models. 
Based on this argument, we have evaluated the constraints on the net baryon-to-entropy ratio $n_B/s$ in this model taking into account reheating processes with different equation of states. 
We have explicitly shown that the present baryon asymmetry 
cannot be generated for $w \leq1/3$, 
which includes the usual inflaton oscillation with a mass term dominated era,  
whatever inflation models are assumed. 
We have found, however,  that, for equation of states $1/3 < w \leq 1$, 
the net baryon-to-entropy ratio can be 
comparable to the observed value with a relatively low reheating temperature
since the entropy production in this scenario is less efficient and the effect is 
larger for lower reheating temperature, 
which is complementary to the vanilla leptogenesis~\cite{Fukugita:1986hr}.   
We have pointed out that the kination domination scenario with $w=1$ is a working example. 
We emphasize that the compared to the usual potential-driven inflation, 
$k$-inflation followed by the kination is more compatible with the scenario since the 
gCS coupling is shift-symmetric and forbids the potential for inflaton. 
Moreover, a larger CP-violation parameter $\Theta$ can be obtained with a smaller gCS coupling
in terms of the ${\cal N}$ parameter. 
Thus, we conclude that our model with $k$-inflation accompanied by gravitational reheating 
can be one of the plausible candidates for the gravitational leptogenesis scenario.

The reheating in the kination scenario is not trivial since the inflaton does not oscillate and 
inflaton decay does not take place. 
However, we have shown that the gravitational reheating with the decay of massive scalar field 
conformally coupled to gravity produced gravitationally at the end of 
inflation~\cite{Hashiba:2018iff}  can work well in this leptogenesis scenario. 
Indeed, we have confirmed that reheating temperature can be small enough to generate 
observed baryon-to-entropy ratio within a feasible range of the parameters. 

The difficulty in explaining the baryon asymmetry of the Universe today by the gravitational
leptogenesis scenario lies in the potential appearance of ghost-like behavior in the high-$k$
mode in the gravitational waves. 
This results in the impossibility of generating huge chiral gravitational waves. 
Indeed, it has been argued that by introducing the Gauss-Bonnet term additionally~\cite{Satoh:2007gn,Satoh:2008ck,Satoh:2010ep, Koh:2014bka,Koh:2016abf,Kawai:2017kqt,Koh:2018qcy}
the amplitude of the chiral gravitational waves can be enhanced. 
Even in this case one should be careful for  the appearance of ghost-like modes, 
on which we here do not explore in depth. 
However, their appearance does not directly mean the catastrophe of the model
but simply means the breakdown of the effective theory. 
We do not rule out the possibility that this singularity is resolved by a healthy 
UV physics~\cite{Lyth:2005jf}, 
and hence our conclusion is a conservative one. 
But without a concrete example, we cannot give any reliable predictions. 

The leptogenesis scenario studied in this paper has a similarity to the baryogenesis 
from the helical hypermagnetic fields generated by the pseudoscalar inflation with 
the $\phi F {\tilde F}$ coupling~\cite{Domcke:2018eki,Domcke:2019mnd} (See also Refs.~\cite{Anber:2015yca,Cado:2016kdp,Jimenez:2017cdr}). 
The leptogenesis in the present case, indeed, corresponds to the particle production through the chiral anomaly 
in the gauge field case, which can change the efficiency of the 
gauge field production~\cite{Domcke:2018eki,Domcke:2018gfr,Domcke:2019mnd}. 
In our study we do not take into account the back reaction on the gravitational wave production 
from the particle production. 
We expect that it does not give a strong effect since the chiral gravitational wave generation is 
a weaker effect than the gauge field production.

\vskip 1cm
\noindent
{\large\bf Acknowledgements}\\
We are grateful to Soichiro Hashiba for useful discussion. 
The work of KK was supported by JSPS KAKENHI, Grant-in-Aid for Scientific Research JP19K03842 
and Grant-in-Aid  for Scientific Research on Innovative Areas 19H04610. 
JK is supported by a research program of the Leading Graduate Course for Frontiers of Mathematical Sciences and Physics (FMSP). 
YY is supported by JSPS KAKENHI, Grant-in-Aid for JSPS Fellows JP19J00494. JY is supported by JSPS KAKENHI, Grant-in-Aid for Sci- entific Research 15H02082 and Grant-in-Aid for Scientific Research on Innovative Areas 15H05888.

\bibliographystyle{JHEP}
\bibliography{ref}
\end{document}